\documentclass[lettersize,journal]{IEEEtran}
\usepackage{bm}
\usepackage{cite} 
\usepackage{amssymb}
\usepackage{graphicx}
\usepackage{subfigure}
\usepackage{algorithmic}
\usepackage{graphicx}
\usepackage{graphics}
\usepackage{textcomp}
\usepackage{subfigure}
\usepackage{tabularx}
\usepackage{times}
\usepackage{multirow}
\usepackage{makecell}
\usepackage{bm}
\usepackage{latexsym}
\usepackage{booktabs}
\usepackage{threeparttable}
\usepackage{epsf}
\usepackage[ruled]{algorithm2e}
\usepackage{color,colortbl}
\usepackage{float}
\usepackage[table,xcdraw]{xcolor}
\usepackage{booktabs}
\begin{document}
	
	\title{Goal-Oriented Integration of Sensing, Communication, Computing, and Control for Mission-Critical Internet-of-Things}
	\author{Jie Cao,~\IEEEmembership{Member,~IEEE},	
		Ernest Kurniawan,~\IEEEmembership{Senior Member,~IEEE},	
		Amnart Boonkajay,~\IEEEmembership{Member,~IEEE},\\
		Sumei Sun,~\IEEEmembership{Fellow,~IEEE},
		Petar Popovski,~\IEEEmembership{Fellow,~IEEE}, Xu Zhu,~\IEEEmembership{Senior Member,~IEEE}	
	\thanks{Jie Cao and Xu Zhu 	are with the School of Electronic and Information Engineering, Harbin Institute of Technology, Shenzhen
		518055, China (Corresponding author: Jie Cao, e-mail: caojhitsz@ieee.org).}\\
	\thanks{ Ernest Kurniawan,  Boonkajay Amnart and Sumei Sun are with the Institute of Infocomm Research, Agency for Science, Technology and Research, Singapore 138632.}
	\thanks{Petar Popovski is with the Department of Electronic Systems, Aalborg	University, Danish.}
	}
	

	
	
	\maketitle
	
	\begin{abstract}
		Driven by the development goal of network paradigm and demand for various functions  in the sixth-generation (6G) mission-critical Internet-of-Things (MC-IoT), we foresee a goal-oriented integration of sensing, communication, computing, and control  (GIS3C) in this paper. 
		We first provide an overview of the tasks, requirements, and challenges of  MC-IoT.  
		Then we introduce an end-to-end GIS3C architecture, in which goal-oriented communication is leveraged to bridge and empower sensing, communication, control, and computing functionalities. 
		By revealing the interplay among multiple subsystems in terms of key performance indicators and parameters,  this paper introduces unified metrics, \emph{i.e.}, task completion effectiveness and cost, to facilitate S3C co-design in MC-IoT. 
		The preliminary results demonstrate the benefits of GIS3C in improving task completion effectiveness while reducing costs.
		We also identify and highlight the gaps and challenges in applying GIS3C in the future 6G networks.
		

	\end{abstract}
	
	\begin{IEEEkeywords}
		 Goal-oriented communication, integration, mission-critical Internet-of-Things.
	\end{IEEEkeywords}
	
	\section{Introduction}
	Mission-critical Internet-of-Things (MC-IoT)  have recently gathered a significant attention, showing the potential to revolutionize many emerging applications such as industrial metaverse and intelligent transportation\cite{ref1}. 
	However, the fast growth of completely automated, highly dynamic, and fully intelligent IoT networks   is likely to exceed the capability of the latest fifth-generation (5G) wireless systems\cite{ref2}.
	Future sixth-generation (6G) MC-IoT requires further advances in current 5G systems for  improving the ability to support various functions\cite{ref3}.
	As shown in Fig. 1 and Table I, an intelligent task typically involves a set of complex  processes.
	It has been a challenge to  meet the stringent and heterogeneous requirements of multiple subsystems simultaneously, due to the complex interdependence among various subsystems with limited network and hardware resources. 


	Both academia and industry have recently shown considerable interest in researching extreme ultra-reliable and low-latency communication, to facilitate the development of 6G MC-IoT\cite{ref2}. 
	Mobile edge computing (MEC) is able to reduce transmission latency, and then  enable more reliable and secure communications.
	Big data with artificial intelligence  (AI) enables accurate prediction and reasonable decision-making to improve transmission efficiency  and intelligence. 
	Moreover, there has been substantial independent research dedicated to enhancing the performance of sensing, control and, computing, with numerous innovative methods being proposed\cite{ref4,ref3gpp}.

	\begin{figure}[tbp]
		\centering
		{\includegraphics[height=5.5cm]{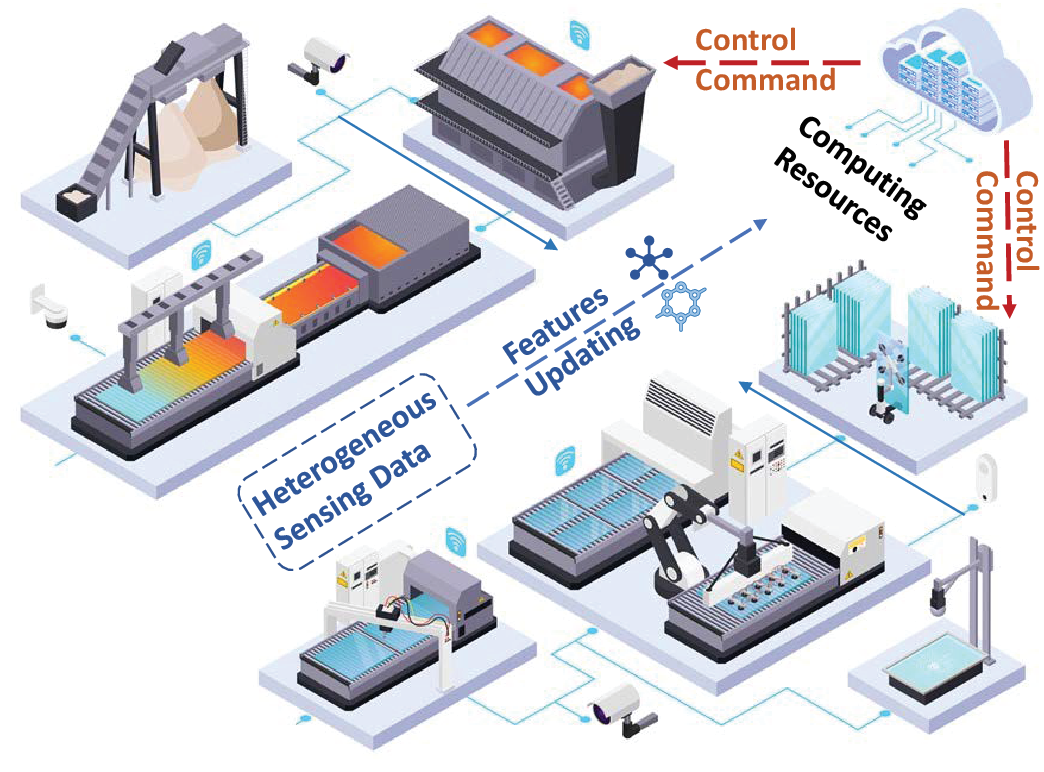}}
		\caption{An Example of 6G MC-IoT: Smart Factory with Sensing, Communication, Control and Computing Functionalities.}
	\end{figure}
	
	However,  rather than struggling to overcome physical  and technical limits to implement  these underlying functionalities separately,  sensing-communication-control-computing (S3C) co-design  appears to be a more  promising solution\cite{ref5}.
	For example, the performance of sensing can be influenced by the allocation of computing resources, which  also has an impact on data size and type in transmission. Likewise, control performance may be limited by wireless fading channels and insufficient computing power. 
	Recently, there has been a widespread and comprehensive investigation into the domain of integrated sensing, communication, and computing\cite{ref6},  showcasing the benefits of co-design. 
	Though these studies have inspired the investigation of S3C co-design,   there remain several unresolved challenges in its realization:
	\textit{(a) a huge amount of heterogeneous data leads to high processing burden and low transmission efficiency, (b) different parameters and lack of unified metrics  hinder system-level global optimization.}  
	
	
	\begin{table*}[t]
		\centering
		\caption{Tasks and Requirements of 6G MC-IoT\cite{ref4,ref3gpp,ref5,ref6}}
		\renewcommand\arraystretch{1.1}
		\begin{tabular}{
				>{\columncolor[HTML]{EFEFEF}}c c
				>{\columncolor[HTML]{EFEFEF}}c c
				>{\columncolor[HTML]{EFEFEF}}c }
			\toprule
			\textbf{Task}                                           & \textbf{Sub-task} & \textbf{Use cases}                              & \textbf{Parameter/Variable}                                                                                                            & \textbf{KPI}                                                                                                                                         \\ \midrule
			\cellcolor[HTML]{EFEFEF}                                & Monitoring        & Road status monitoring in smart vehicles       &                                                                                                                               & \cellcolor[HTML]{EFEFEF}                                                                                                                             \\ \cline{2-3}
			\cellcolor[HTML]{EFEFEF}                                & Detection         & Fault detection in smart grid                  &                                                                                                                               & \cellcolor[HTML]{EFEFEF}                                                                                                                             \\ \cline{2-3}
			\multirow{-3}{*}{\cellcolor[HTML]{EFEFEF}Sensing}       & Localization      & localization and tracking in smart factory & \multirow{-3}{*}{\begin{tabular}[c]{@{}c@{}}Sampling type\\ Sampling rate\\ Sampling duration\end{tabular}}                   & \multirow{-3}{*}{\cellcolor[HTML]{EFEFEF}\begin{tabular}[c]{@{}c@{}}Monitoring error\\ Detection accuracy\\ Localization resolution\end{tabular}}    \\ \midrule
			\cellcolor[HTML]{EFEFEF}                                & Reliability       & Communication between vehicles                 &                                                                                                                               & \cellcolor[HTML]{EFEFEF}                                                                                                                             \\ \cline{2-3}
			\cellcolor[HTML]{EFEFEF}                                & Effectiveness     & Energy status updating in smart grid           &                                                                                                                               & \cellcolor[HTML]{EFEFEF}                                                                                                                             \\ \cline{2-3}
			\multirow{-3}{*}{\cellcolor[HTML]{EFEFEF}Communication} & Security          & Interaction between robots and plants          & \multirow{-3}{*}{\begin{tabular}[c]{@{}c@{}}Data size / Channel use\\ Power / Bandwidth\\ Access control / Scheduling\end{tabular}} & \multirow{-3}{*}{\cellcolor[HTML]{EFEFEF}\begin{tabular}[c]{@{}c@{}} Error rate \\ Throughput / Latency \\ Security throughput\end{tabular}}              \\ \midrule
			\cellcolor[HTML]{EFEFEF}                                & Stabilization     & Automotive maneuvering and navigation          &                                                                                                                               & \cellcolor[HTML]{EFEFEF}                                                                                                                             \\ \cline{2-3}
			\cellcolor[HTML]{EFEFEF}                                & Tracking          & Smart grid precise load control                &                                                                                                                               & \cellcolor[HTML]{EFEFEF}                                                                                                                             \\ \cline{2-3}
			\multirow{-3}{*}{\cellcolor[HTML]{EFEFEF}Control}       & Constraints       & Remote control of assembly robots              & \multirow{-3}{*}{\begin{tabular}[c]{@{}c@{}}Control type and law\\ Control period\\ Control limits\end{tabular}}                  & \multirow{-3}{*}{\cellcolor[HTML]{EFEFEF}\begin{tabular}[c]{@{}c@{}}Stability margin\\ Steady state error\\  Overshoot\end{tabular}}        \\ \midrule
			\cellcolor[HTML]{EFEFEF}                                & Training          & MEC assisted autonomous driving     &                                                                                                                               & \cellcolor[HTML]{EFEFEF}                                                                                                                             \\ \cline{2-3}
			\cellcolor[HTML]{EFEFEF}                                & Inference         & Load forecasting in smart grid                 &                                                                                                                               & \cellcolor[HTML]{EFEFEF}                                                                                                                             \\ \cline{2-3}
			\multirow{-3}{*}{\cellcolor[HTML]{EFEFEF}Computing}     & Decision-making   & Predictive control in smart factory            & \multirow{-3}{*}{\begin{tabular}[c]{@{}c@{}}Task assignment\\ Computing resource \\ Learning network\end{tabular}}            & \multirow{-3}{*}{\cellcolor[HTML]{EFEFEF}\begin{tabular}[c]{@{}c@{}}Resource utilization\\ Service response time\\ Service reliability\end{tabular}} \\ \bottomrule
		\end{tabular}
	\end{table*}

	{The quality of 6G mission-critical services is normally characterized by
		the efficiency of completing specific tasks, instead of conventional task-agnostic performance metrics like latency, reliability, and throughput\cite{ref8}.
		Also, inadequate considerations of packet contents and receiver tasks lead to a large number of useless packets being transmitted, resulting in wasted resources and hindering S3C co-design. 
		However, most of the current systems are designed to recover data accurately 	while ignoring the contents/semantics of transmitted packets or their impacts on the receiver.
		Meanwhile, the existing communication technologies
		have nearly approached the physical-layer capacity limit\cite{ref9_1}. Therefore, there is an urgent need to develop a new communication paradigm.
		
		{In this regard, adopting goal-oriented  communication as a bridge between multiple
		functionalities to communicate and exchange desired information can effectively enhance the successful execution of tasks. 
		We believe it is timely to study the goal-oriented integration of sensing, communication, computing, and control (GIS3C)  in 6G MC-IoT\cite{ref9}.
		GIS3C aims at extracting and transmitting the relevant semantic features needed to make the receiver accomplish a goal with the desired effectiveness\cite{ref9_1}.
		This can help to reduce the amount of data and clarify the coupling between multiple subsystems.
		However, GIS3C is still in its infancy and there are a large number of open issues, such as \textit{(a) How  does GIS3C empower MC-IoT and facilitate the co-design of S3C? (b) Is there any unified system-level  metric  to break down the barriers among different subsystems?}
		We are aware that GIS3C can be studied from many perspectives, such as AI-based approaches. In this paper, we focus more on the information flow and data transmission of GIS3C in 6G MC-IoT.
		The main contributions are as follows.

	\begin{itemize}
		\item We first overview the tasks, requirements, and metrics of 6G MC-IoT, from the perspective of data transmission. Also, the challenges of realizing S3C co-design in 6G MC-IoT  are summarized.
		\item 
		We introduce an end-to-end (E2E) GIS3C architecture, based on which environment-aware sensing, semantic communication, context-aware control, and situation-aware computing are analyzed. 
		\item 
		We provide a comprehensive analysis to  reveal	the interplay among multiple subsystems in terms of key performance indicators (KPIs) and parameters. We illustrate how GIS3C can be used to facilitate S3C co-design, and introduce task completion effectiveness and cost as the unified metrics. Furthermore,  preliminary results are provided to verify the effectiveness of the introduced GIS3C method.
	\end{itemize}
	
	The rest of this article is organized as follows.
	Section II first
	reviews  the tasks, requirements and challenges of the integrated system in 6G MC-IoT. 
	Sections III and IV illustrate how 
	{GIS3C} can be used to empower sensing, communication, control and computing, as well as to facilitate  their integration. Conclusion and future work are presented in Section V.

	\section{Tasks and Challenges of S3C for  6G MC-IoT}
	To clarify the coupling between multiple subsystems, we first specify their tasks and requirements in Subsection II-A. The challenges of realizing S3C co-design in 6G MC-IoT are summarized in Subsection II-B.


	
	\vspace{-2pt}
	\subsection{Tasks and Requirements of 6G MC-IoT}
	
	Accurate and timely sensing  is the basis of 6G MC-IoT.
	Multiple smart devices are installed  to collect environmental parameters and observe physical processes, based on which  multiple sub-tasks including monitoring, detection, and localization are completed, as summarized in Table I.  
	For the monitoring task, multiple parameters, \emph{e.g.}, speed, direction, and position,   are required to be monitored accurately, which is evaluated by the monitoring error.
	Also, such measurements provide holistic supervision and  real-time alerts for abnormal conditions\cite{ref2}, where the detection accuracy is the critical KPI and can be improved with a larger sampling rate. 
	Precise positioning of mobile devices is also one of the important sub-tasks of sensing, where a higher localization resolution is obtained by a larger sampling duration.
	
	
	%
	
	{Timely and reliable transmission is the core of implementing 6G MC-IoT.}
	Reliable transmission with a low error rate ensures that the packets can be transmitted successfully, by optimizing data size and channel use. 
	With power and bandwidth allocation, effective transmission with low latency and high throughput can enable massive amounts of status information to be updated in real time.
	For example, extremely reliable communication with negligible latency is necessary to support autonomous and safe driving.  This is supplemented by the need for a secure transmission. 
	

	{Reliable and automatic control is the key to successful task execution in 6G MC-IoT.} 
	Most of control tasks are aimed at the stable operation of the equipment, where the stability margin is designed to be maximized.
	By designing the control law and period, the steady-state error is minimized  for automatic tracking and the overshoot is minimized to ensure that the control task is completed within the given constraints. 
	For example,  load  frequency  control   shares the burden  of power regulation  in the interconnected  power  system. It can  ensure that the grid operates in a stable frequency range while avoiding frequency overshooting, by adjusting the generator to allow power generation to cope with load fluctuations.
	

	{Efficient computing provides the critical support for 6G MC-IoT.}
	MEC can be adopted for training based on collected data, which can assist in understanding situations such as road condition analysis in autonomous driving.
	In addition,  inference and decision-making cannot be accomplished without computation. For example,   load forecasting in smart grid  and predictive control in smart factory require massive amounts of computing resources.
	In  such tasks, multiple KPIs including resource utilization, service response time and reliability are adopted to evaluate the effectiveness of computing.

	\subsection{Challenges of Realizing S3C Co-Design in  6G MC-IoT }
	Even though the rapid development of 5G and the introduction of several emerging  technologies have made mission-critical services possible, the following challenges remain unexplored in realizing S3C co-design in 6G MC-IoT.	
	
	{\textbf{Huge Amounts of Data:}}
	To constantly track the state of the environment, a great number of connected sensors and devices are employed, which generates large amounts of data.  Hence, it can be a challenging task to store, process, and analyze big data while ensuring low latency and high efficiency.
	To alleviate the burden,  MEC has been widely used  to mine and aggregate data in a decentralized manner, which brings  the  possibility to enable low latency transmission.
	However, a large amount of data is still discarded even with MEC, resulting in wasted resources, as much of the information is not useful to accomplish a certain task such as fault detection and automatic control. 

	\textbf{Multiple sources with heterogeneous traffic:}
	The 6G MC-IoT paradigm encompasses a significant portion of big data, which is characterized by heterogeneity, manifested in high dimensionality and diverse forms of expression, as shown in Table I.
	This situation poses a major challenge for data management due to the introduction of multiple sources with different attributes.
	Additionally, the communication system must be capable of accommodating heterogeneous data and diverse types of traffic. Specifically, the coexistence of short and long packets, as well as the presence of both delay-sensitive and tolerant data, necessitates the development of a robust and dependable communication infrastructure.
	Different data types and structures for sensing, communication, and control complicate integration design.

	\textbf{Coupled Multiple Functionalities:}	
	As highlighted in Subsection II-A,  the 6G MC-IoT  comprises interconnected subsystems, including sensing, communication, control, and computing, which exhibit intricate interdependencies. 
	Therefore,  to design a comprehensive and fully integrated system, a holistic approach is required. However, the dynamics and interactions between these subsystems remain ambiguous and the impact of different parameters on system performance needs to be further investigated. The absence of standardized protocols and unified metrics poses a substantial challenge when it comes to integrating multiple subsystems.

	\begin{figure*}[tbp]
		\centering
		{\includegraphics[height=9.2cm]{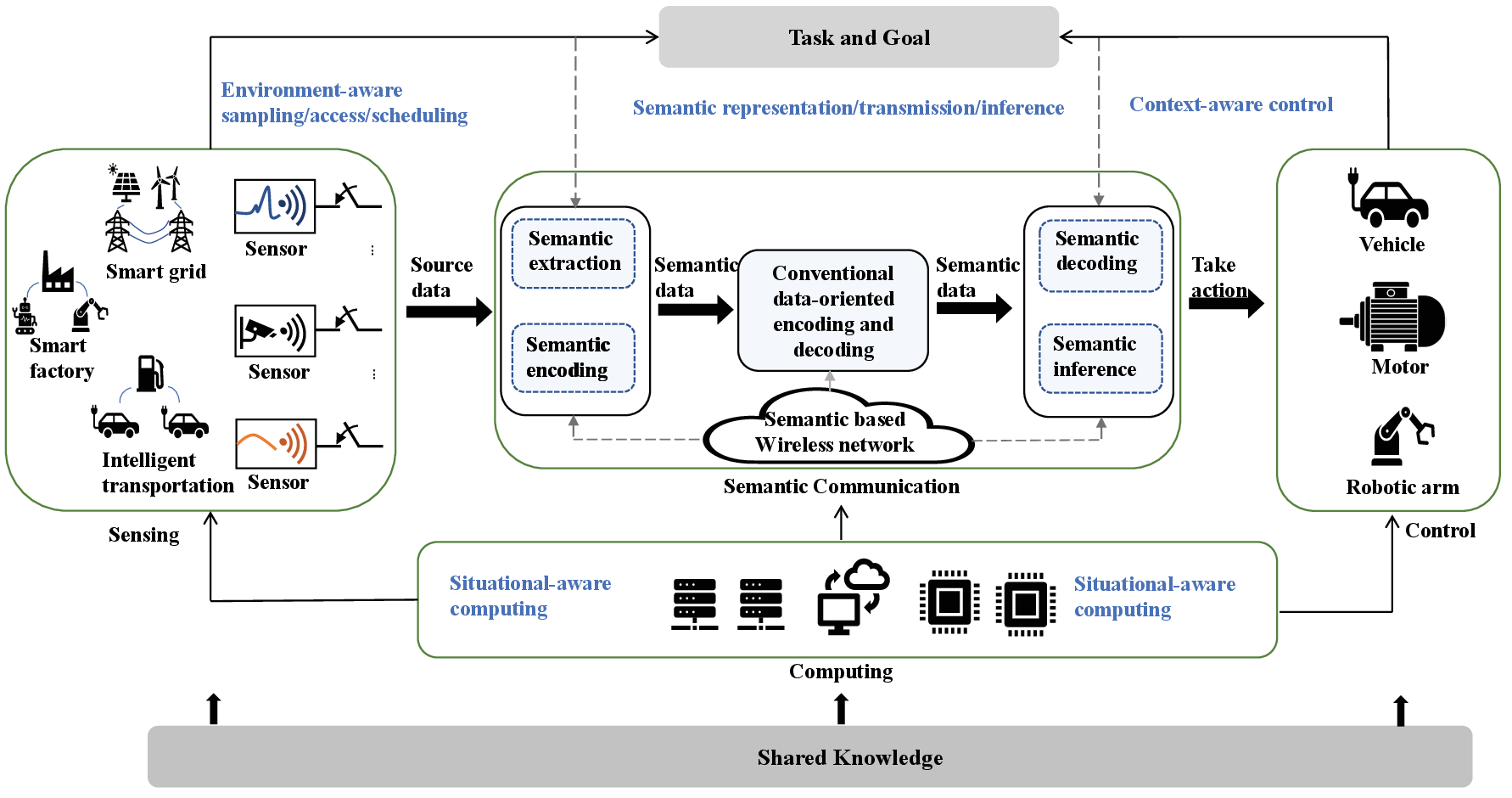}}
		\caption{E2E {GIS3C} Architecture in 6G MC-IoT.}
	\end{figure*}
	\section{E2E GIS3C for 6G MC-IoT}
	
	
	In this section, we propose an E2E GIS3C architecture in 6G MC-IoT, based on which {GIS3C}-empowered multiple subsystems are illustrated.

	\subsection{GIS3C in 6G MC-IoT}
	
	As shown in Fig. 2, semantic communication (SemCom), focusing on transmitting symbols convey the semantic and effectiveness level of goal-oriented communication, serves as a bridge among sensing, control, and computing subsystems. 
	Based on pioneering work done 
	by Weaver\cite{ref8}, GIS3C  is designed to extract and transmit the relevant information needed to make the receiver accomplish a goal with the desired  effectiveness.
	Furthermore, {GIS3C} is expected to 
	empower multiple subsystems by making them more ``understandable'' and ``communicable'' based on the common task and shared knowledge, as elaborated in the following subsections. 

	\subsection{Environment-aware Sensing}
	In 6G MC-IoT, multiple sensors monitor the physical process based on the given task/goal and environment. 
	Traditional sensing can be empowered by GIS3C for environment-aware sensing.
	Unlike the traditional one that samples and transmits all packets, \textbf{only important, relevant, and urgent information is sampled  in environment-aware sensing, which is able to reduce the amount of data. Also, these data are assigned more resources with higher priority.} 
	To achieve this, the foremost and crucial thing is to characterize the semantic attributes and guide the design of sampling and scheduling\cite{ref11}. 
	However, determining semantic attributes of sensing packets based on the task and common knowledge is still challenging. 
	In this paper, we  focus on 
	various semantic metrics from the effectiveness level, as discussed below.
	\subsubsection{Time-Based Metric}	
	Age of Information (AoI) is a measure of freshness for an information flow, a typical time-based semantic metric, defined as the time elapsed since the latest successfully received packet  was generated at the source\cite{ref11_1}.
	Moreover, the value of information   and age of loop   were proposed to evaluate the impact of AoI on different applications and  characterize the freshness of information in the  closed-loop system.
	However, AoI-based metrics imply that the freshest packet has the most valuable information, which is not applicable to all  scenarios.
	
	\subsubsection{Error-Based Metric}
	
	Mean square error (MSE) is a well-known metric that characterizes the accuracy aspect of information, which serves as one of the important semantic attributes.
	The importance of information is explicitly defined by the real-time square error between transceivers, based on which the sampling is triggered by the state error\cite{ref8}.

	\subsubsection{Goal-Based Metric}
	Combining time-based metrics with
	error-based metrics, 
	age of incorrect information (AoII) characterizes the impact of the prolongation of one inaccurate state on semantic recovery\cite{ref8}. 
	Also, a context-based metric, named as urgency of information (UoI), has been proposed to measure the importance of
	the non-uniform context-dependence of state information\cite{ref11}.
	Furthermore,  considering the specific control task, a goal-oriented sampling and scheduling policy was proposed for the minimization of the actuation error\cite{ref12}.

	

	\begin{figure*}[tbp]
		\centering
		{\includegraphics[height=5cm]{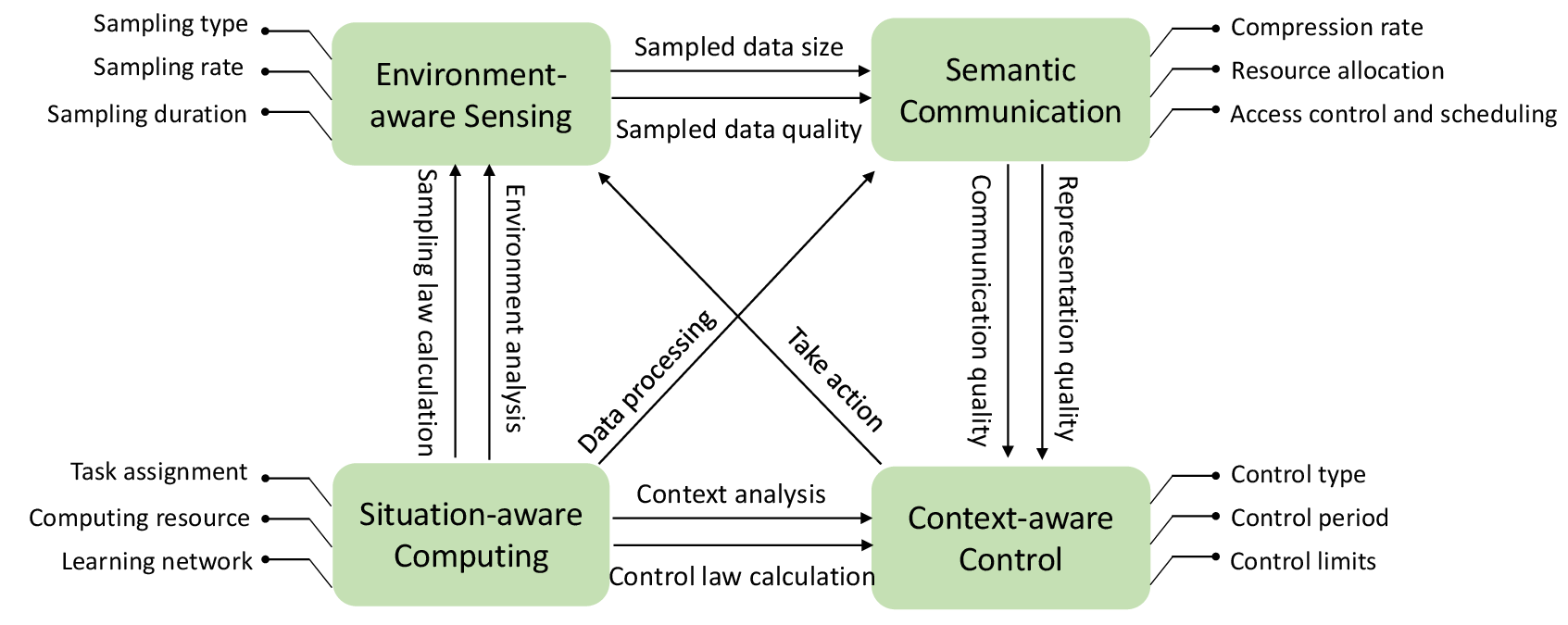}}
		\caption{Interplay Among Multiple Subsystems in {GIS3C}-assisted 6G MC-IoT.}\label{fig_4}
	\end{figure*}
	
	%
	
	\vspace{-10pt}
	\subsection{Semantic Communication}
	Usually, the semantic features are obtained by  extracting and encoding the transmitted source data, and then filtered and compressed  according to the  common task and shared knowledge.
	After that, semantic data is transmitted through traditional data-oriented encoding and decoding over  wireless networks\cite{ref9_1}. \textbf{{GIS3C} allows only the transmission of semantic features of interest to the task, rather than raw data, which alleviates bandwidth pressure and reduces the redundant data.}
	However, semantic extraction/filtering for various data types with dynamic tasks and network states is challenging.  
	The key techniques are discussed as follows.
	
	\subsubsection{Semantic Extraction and Filtering}
	Several semantic extraction technologies have been explored in the field of computer science.
	For different data types such as text, audio, and image,  natural language processing, speech signal processing, and computer vision have attracted extensive attention and have been thoroughly researched\cite{ref9}. Thanks to the development of big data and computing ability, several AI-based semantic extractions such as deep learning-based, reinforcement learning-based and knowledge base-based techniques have been investigated for different scenarios.


	\subsubsection{Semantic Encoding}
	
	Besides physical turbulence 	and noise (\emph{e.g.}, Gaussian noise and multi-path fading),
	semantic noise also involves semantic mismatch, ambiguity, and interpretation errors. Consequently, beyond simple data compression, {GIS3C} strives to effectively combat the semantic  	noise and transmit the semantic meaning  with adequate encoding and decoding  	schemes.
	Considering the task at the receiver, it is important but  challenging to quantify semantic information and derive semantic capacity.
	Furthermore, the robustness of {GIS3C} can be improved by  integrating the channel information  into semantic encoding/decoding, or by considering source and channel encoding/decoding together, which has attracted a lot of attention.

	
	
	
	\vspace{-10pt}
	\subsection{Context-Aware Control}
	Based on the decoded semantic meaning, the controller applies reasoning methods for obtaining high-level contextual information and can proactively correct transmission-induced errors. 
	In contrast, conventional controllers can only compute control commands based on received packets.
	\textbf{Context-aware control 	allows the prediction of outcomes through  the complex awareness of
		the process}, thus exploiting all data and emerging  an optimal solution based on advanced computing.
	Specifically, the accuracy of context awareness depends on the context modeling and reasoning, which is challenging for coupled systems with heterogeneous traffic and dynamic tasks.  The key techniques are discussed as follows.
	\subsubsection{Modeling}
	Generally, different data types such as sensing, communication, and computing are typically presented in different formats that might not be readily understandable to the user or device.
	Based on the specific task and common knowledge, context modeling is adopted to represent these data into meaningful terms, which are expected to be simplicity, reusability, and scalability.
	Context modeling is achieved through a variety of approaches, such as graphical models, logic-based and ontology-based models\cite{ref18}, which are suitable for different scenarios and requirements.
	\subsubsection{Reasoning}
	Reasoning or the evaluation of context involves extracting new knowledge from the modeled data of the available context. This step can be divided into three phases: a) pre-processing of context data to eliminate inaccurate values; b) fusion of sensor data to generate more precise information; c) context inference to obtain new context information from lower-level context sources.
	Furthermore, techniques for processing the contextual available input can be classified as learning or inference, such as rule-based and fuzzy logic-based methods for the classification task, as well as learning-based methods for the clustering task\cite{ref18}.

	%

	\vspace{-10pt}
	\subsection{Situation-Aware Computing}
	A strong computational base enables intelligent sensing, semantic communication,  and context-aware control. 
	By allocating computing resources in data sampling, features representation, and inference as well as control command computing dynamically,  resource utilization can be improved. In contrast, the pre-allocation of fixed computing resources leads to inefficiencies. 
	\textbf{Situation-aware computing is able to offload tasks and allocate resources dynamically and intelligently based on the tasks.}
	However, establishment of knowledge bases and adaptive resources allocation are challenging,   as discussed below.
	
	\subsubsection{Task Offloading and Allocation}
	Task offloading is the process of transferring computation-intensive tasks to a set of remote computing machines (\emph{e.g.}, cloud or edge servers) that can process
	the tasks. 
	Specifically, task offloading is adopted based on the requirements and common knowledge by the following three steps: 
	a) task priority assignment, b) redundant task elimination, and
	c) task scheduling.
	An efficient task offloading strategy can significantly  reduce the latency and energy consumption while improving  task execution efficiency. 
	
	\subsubsection{Resource Provisioning}
	Determining sufficient computational resources for performing each task is challenging due to the complex coupling among S3C.
	For example, since the  computation and communication compete against each other  for the shared time resource, a higher computing delay is  leading to less time budget for communication, and therefore  more transmission errors, \emph{i.e.}, degraded communication dependability.  On the other hand, the reduction in computation  error tends to reduce the value error, and therefore increase  the control dependability. 
	Furthermore, with a longer learning process and higher central processing unit (CPU) rate, a more compressed and accurate semantic feature with lower inference error can be obtained.

	
	%
	
	\begin{figure*}[tbp]
		\centering
		{\includegraphics[height=7.4cm]{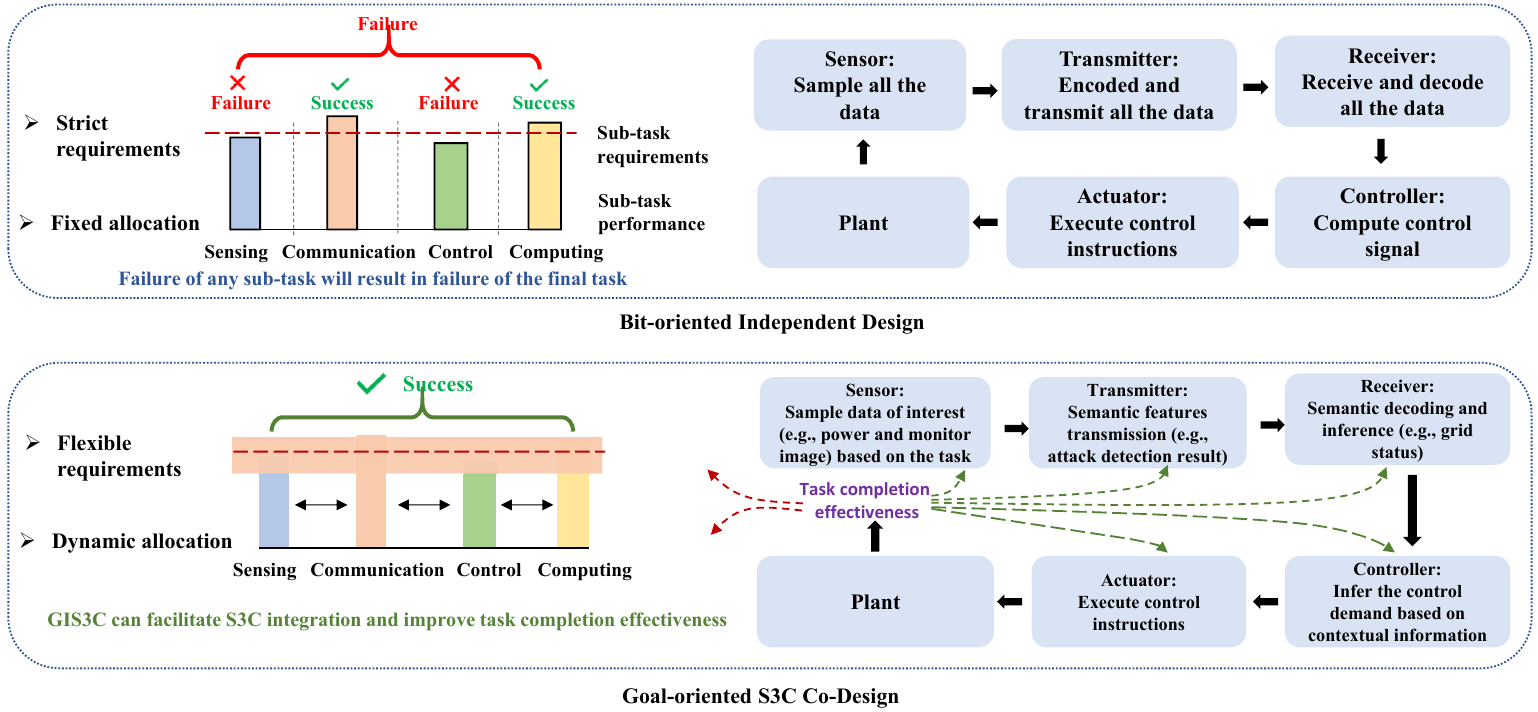}}
		\caption{Comparison Between Bit-oriented Independent Design (the upper part) and Goal-oriented S3C Co-Design (the bottom part).}\label{fig_3}
	\end{figure*}
	%
	
	\section{GIS3C-Assisted 6G MC-IoT}
	
	In the last section, we have shown that multiple subsystems can be enhanced by {GIS3C}. However, the independent design of each subsystem still results in low resource utilization. In this section, we first investigate the interplay among different subsystems, and then illustrate how {GIS3C} can be used in 6G MC-IoT and provide a use case to verify our method. 
	
	\vspace{-10pt}
	\subsection{Interplay among S3C}

	As shown in Fig. 3, sensing, communication, control and computing are tightly coupled with each other by different parameters and KPIs.
	
	\textbf{Environment-aware sensing is performed based on the analyzed environment states and control actions, which provides  high-quality  sampled data for transmission.}
	The sampled data size depends on the sampling type (\emph{e.g.}, event-triggered or time-triggered), sampling rate and sampling duration, which induces the sampling cost $\mathcal{C}_{sa}$ in terms of energy or computing resources. 
	Also, these parameters affect the sensing performance $\mathcal{J}_{sa}$, such as monitoring error and detection accuracy.
	For example, a larger monitoring duration can obtain a smaller detection error but requires more energy consumption. 
	Sampled data size and sensing performance in turn affect the performance and overhead of semantic representation.
	
	\textbf{Semantics extraction/filtering are accomplished through computation, and transmitted over wireless networks for control command calculation.}
	The sampled source data is extracted and encoded as semantic data for transmission. The receiver decodes the semantic features and infers their meaning.
	The effectiveness of feature extraction depends on the allocation of computing resources $\mathcal{C}_{cp}$. The compression ratio also affects the communication overhead $\mathcal{C}_{cm}$  the performance of semantic inference $\mathcal{J}_{cm}$.
	Furthermore, resource allocation scheduling over wireless networks influences the transmission performance of semantic data.
	A lower compression ratio with limited resource allocation can save the communication cost $\mathcal{C}_{cm}$ but incur more inference errors.
	The semantic communication performance includes the transmission error and inference error, which is affected by the allocated communication and computing resources, as well as the optimized communication parameters.
	
	\textbf{Control commands are inferred and reasoned from decoded semantic information based on context analysis and common knowledge.}
	The controller computes the control demand based on the inferred semantic meaning and context,  and then provides guidance for taking action. 
	On one hand,   making decisions based on the inferred information and context consumes a lot of computing resources $\mathcal{C}_{cp}$ and control resources $\mathcal{C}_{cn}$. Its  correctness is also influenced by the allocated resources and the semantic inference performance $\mathcal{J}_{cm}$.
	On the other hand, packet loss and transmission delay may lead to inaccurate operations, resulting in worse control performance $\mathcal{J}_{cn}$ such as system downtime and security risks.

	\subsection{Goal-Oriented System-Level Performance Metric}

	Based on the coupling analysis among multiple subsystems, this subsection compares the proposed {GIS3C} scheme with the  bit-oriented independent design method, as shown in Fig. 4. 
	Particularly, we introduce the task completion effectiveness as the main KPI, \emph{i.e.}, $\mathcal{J}_{sys}=\chi(\mathcal{J}_{sa},\mathcal{J}_{cm},\mathcal{J}_{cn},\mathcal{J}_{cp})$, which measures the probability and satisfaction that goals are met or tasks are completed. 
	The system-level task completion effectiveness depends on the performance of multiple subsystems with the mapping function $\chi(\cdot)$.
	Similarly, the system-level task cost can be obtained by converting all the costs into the same domain, \emph{i.e.}, $\mathcal{C}_{sys}=c_{sa}(\mathcal{C}_{sa})+c_{cm}(\mathcal{C}_{cm})+c_{cn}(\mathcal{C}_{cn})+\mathcal{C}_{cp}$, where $c(\cdot)$ is the transformation function for different domain costs. The mapping functions $\chi(\cdot)$ and $c(\cdot)$ can be derived based on the analysis in Subsection IV-A and specific tasks. 
	
	With the system-level task completion effectiveness and cost, multiple subsystems can be designed jointly with dynamic resource allocations and flexible optimizations.
	Furthermore, this can help to reduce the amount of data and integrate multiple subsystems with heterogeneous traffic by representing different information as a unified way based on the task and common knowledge.
	In contrast,  in the conventional independent design, resource allocations for each subsystem must be determined in advance based on requirements, in which any failure of subsystems may result in system downtime. 
	With the assistance of {GIS3C},   the task completion effectiveness can be improved with less cost by collaborating multiple subsystems.
	For example, if the controller receives/decodes contradictory information from different sources, it can analyze these information globally and compare these with historical information, and then infer the correct control command based on contextual information.

	%


	\subsection{ Use Case}
	In this subsection, we  consider the load frequency control (LFC) system in smart grid as a case study. 
	The goal of the considered LFC system is to maintain the balance between power load and generation at minimal cost\cite{globecom23_CJ}.  
	To ensure the stability of power systems, multiple sensors are employed to monitor the frequency deviation and detect cyber-attacks. Then the controller computes control commands, guiding  the actuator (\emph{e.g.}, engine or motor) to adjust power generation to ensure grid stability. Three methods are introduced and compared as follows.
	
	\textbf{Bit-oriented independent design:}
	Pre-allocate fixed resources to each subsystem based on the relationship between sub-tasks and the system-level task, as shown in the upper part of Fig. 4.
	
	\textbf{Bit-oriented co-design:}
	Dynamically allocate resources to each subsystem to maximize their performance with the total cost constraint.
	
	\textbf{Goal-oriented co-design:}
	Analyze the interplay among {GIS3C}-empowered multiple subsystems and their relationship to the system-level task, and then dynamically allocate resources to each subsystem for maximizing the task completion effectiveness with the cost constraint, as shown in the bottom part of Fig. 4.
	In the considered LFC system, resources such as  sensing power, CPU rate, wireless communication resources (\emph{e.g.}, bandwidth or power) affect the system stability in different ways, as analyzed in Subsections IV-A and IV-B.
	
	\begin{figure}[tbp]
		\centering
		{\includegraphics[height=5.3cm]{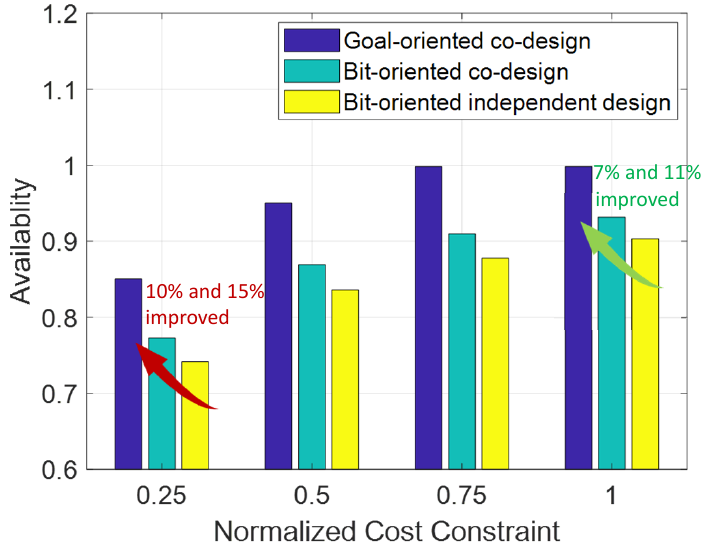}}
		\caption{Performance Comparison Among Different Design Methods in Terms of Task Completion Effectiveness and Costs.}\label{fig_6}
	\end{figure}
	
	Fig. 5 verifies the effectiveness of our proposed {GIS3C} method. 
	Task completion effectiveness is defined as the availability of the grid, which measures the probability that the grid operates within the allowable frequency deviation range.
	Also,  task costs for multiple subsystems are converted and normalized into energy consumption. 
	Compared to the existing methods,  the proposed {goal}-oriented co-design can achieve a higher availability with the same cost constraints, especially with a lower cost constraint (up to 15\% improvement). This is due to that with the assistance of {GIS3C}, resources can be  allocated dynamically and intelligently to different subsystems to assist in accomplishing system-level tasks.

	\section{Conclusions and Future Work}
	
	In this article,  we have introduced the GIS3C for 6G MC-IoT, which sheds light on the development of future 6G networks. The tasks, requirements, and challenges of supporting 6G MC-IoT  have been overviewed. We have provided a comprehensive introduction to E2E {GIS3C} architecture. In particular, environment-aware sensing, semantic communication, context-aware control, and situation-aware computing have been analyzed. Additionally, the interplay among multiple subsystems has been revealed,  based on which a system-level metric has been proposed to facilitate S3C co-design in MC-IoT. 
	
	Although this article has pointed out some possible research opportunities in the {GIS3C}, there are still many gaps and challenges in applying {GIS3C} in 6G MC-IoT. Therefore, in the following, we discuss  some possible research directions for future goal-oriented SemCom and S3C in 6G MC-IoT, such as
	a) Real-time green semantic communication: semantic communication requires more computing resources for feature extractions and model training, which further complicates the power control and energy management. How to balance the energy efficiency and stringent communication performance of the 6G MC-IoT with limited bandwidth provisioning and low transmit power remains unexplored.
	b) Co-existence of heterogeneous networks: for the co-existence of bit-oriented and goal-oriented networks, interaction analysis and optimization are still in their early stage. 
	(c) Semantic native communication: How to implement SemCom in task/goal-unaware systems with the unknown environment requires further investigation. Also, it is worth investigating how to design adaptive (universal) SemCom with implicit semantics in 6G MC-IoT.

	\newpage
	
	
	\vspace{11pt}
	
	
	%

	\vfill
	

\begin{thebibliography}{1}
		\bibliographystyle{IEEEtran}
		%
		
		\bibitem{ref1}
		J. Cao et al., ``Toward Industrial Metaverse: Age of Information, Latency and Reliability of Short-Packet Transmission in 6G,''  {\em IEEE Wireless Communications}, vol. 30, no. 2, pp. 40-47, April 2023.
		
		
		%
		
		\bibitem{ref2}
		S. He, K. Shi, C. Liu, B. Guo, J. Chen and Z. Shi, ``Collaborative Sensing in Internet of Things: A Comprehensive Survey,''  \emph{IEEE Communications Surveys \& Tutorials}, vol. 24, no. 3, pp. 1435-1474, thirdquarter 2022.
		
		
		\bibitem{ref3}
		X. Hou, J. Wang, Z. Fang, Y. Ren, K. -C. Chen and L. Hanzo, ``Edge Intelligence for Mission-Critical 6G Services in Space-Air-Ground Integrated Networks,''  \emph{ IEEE Network}, vol. 36, no. 2, pp. 181-189, April 2022.
		
		\bibitem{ref4}
		P. Park, S. Coleri Ergen, C. Fischione, C. Lu and K. H. Johansson, ``Wireless Network Design for Control Systems: A Survey,''  \emph{IEEE Communications Surveys \& Tutorials}, vol. 20, no. 2, pp. 978-1013, Secondquarter 2018.
		
		\bibitem{ref3gpp}
		\emph{3GPP Technical Report, TR 22.837 V2.0.0}, 	``Feasibility study on integrated sensing and
		communication,'' 15 June 2023.
		
		\bibitem{ref5}
		Adam MM, Zhao L, Wang K, Han Z., ``Beyond 5G Networks: Integration of Communication, Computing, Caching, and Control,'' \emph{arXiv preprint}, doi: arXiv:2212.13141, 2022.
		
		\bibitem{ref6}
		U. Demirhan and A. Alkhateeb, ``Integrated Sensing and Communication for 6G: Ten Key Machine Learning Roles,'' \emph{ IEEE Communications Magazine}, vol. 61, no. 5, pp. 113-119, May 2023.
		
		
		
		
		\bibitem{ref8}
		W. Yang et al., ``Semantic Communications for Future Internet: Fundamentals, Applications, and Challenges,'' {\em IEEE Communications Surveys \& Tutorials}, Early Access, 2022.
		
		
		\bibitem{ref9_1}
		X. Luo, H. -H. Chen and Q. Guo, ``Semantic Communications: Overview, Open Issues, and Future Research Directions,'' \emph{ IEEE Wireless Communications}, vol. 29, no. 1, pp. 210-219, February 2022.
		
		
		\bibitem{ref9}	
		D. Gündüz et al., ``Beyond Transmitting Bits: Context, Semantics, and Task-Oriented Communications,'' {\em IEEE Journal on Selected Areas in Communications}, vol. 41, no. 1, pp. 5-41, Jan. 2023.
		
		
		
		
		
		
		
		
		
		
		
		
		
		
		
		
		
		
		
		
		
		
		
		
		
		
		
		
		
		
		
		
		
		\bibitem{ref11}
		M. Kountouris and N. Pappas, ``Semantics-Empowered Communication for Networked Intelligent Systems,''  {\em IEEE Communications Magazine}, vol. 59, no. 6, pp. 96-102, Jun. 2021.
		
		\bibitem{ref11_1}
		Antzela Kosta; Nikolaos Pappas; Vangelis Angelakis, ``Age of Information: A New Concept, Metric, and Tool,'' {\em Now Foundations and Trends}, 2017.
		
		\bibitem{ref12}
		E. Fountoulakis, N. Pappas and M. Kountouris, ``Goal-Oriented Policies for Cost of Actuation Error Minimization in Wireless Autonomous Systems,''  {\em IEEE Communications Letters}, vol. 27, no. 9, pp. 2323-2327, Sept. 2023.
		
		
		
		\bibitem{ref18}
		R. A. C. Diaz, M. Ghita, D. Copot, I. R. Birs, C. Muresan and C. Ionescu, ``Context Aware Control Systems: An Engineering Applications Perspective,'' {\em IEEE Access}, vol. 8, pp. 215550-215569, 2020.
		
		
		\bibitem{globecom23_CJ}
		J. Cao, E. Kurniawan,  A. Boonkajay and S. Sun, ``Goal-Oriented Scheduling and Control Co-Design in Wireless Networked Control Systems,'' to be published in {\em proc. IEEE Globecom}, Dec. 2023.
		
		
		
		
		
		
		
		
		%
		%
		%
		%
		%
		%
		%
		%
		
	\end{thebibliography}
\end{document}